\documentclass[aps, prd, reprint, twocolumn, superscriptaddress]{revtex4-1}
\usepackage{amsmath}
\usepackage{amssymb}
\usepackage{graphicx}
\usepackage{color}
\usepackage{times}
\usepackage{euscript}
\usepackage{amsthm}
\usepackage{amsfonts}
\usepackage[english]{babel}
\usepackage{epstopdf}
\usepackage{multirow,makecell,array}
\usepackage{mathtools}
\usepackage{float}
\usepackage[dvipsnames]{xcolor}

\begin{document}

\title{Stimulated vacuum emission and photon absorption in strong electromagnetic fields}

\author{I.~A.~Aleksandrov}
\affiliation{Department of Physics, Saint Petersburg State University, Universitetskaya Naberezhnaya 7/9, Saint Petersburg 199034, Russia}
\affiliation{Ioffe Institute, Politekhnicheskaya street 26, Saint Petersburg 194021, Russia}
\author{A.~Di~Piazza}
\affiliation{Max Planck Institute for Nuclear Physics, Saupfercheckweg 1, Heidelberg D-69117, Germany}
\author{G.~Plunien}
\affiliation{Institut f\"ur Theoretische Physik, TU Dresden, Mommsenstrasse 13, Dresden D-01062, Germany}
\author{V.~M.~Shabaev}
\affiliation{Department of Physics, Saint Petersburg State University, Universitetskaya Naberezhnaya 7/9, Saint Petersburg 199034, Russia}

\begin{abstract}
According to quantum electrodynamics (QED), a strong external field can make the vacuum state decay producing electron-positron pairs. Here we investigate emission of soft photons which accompanies a nonperturbative process of pair production. Our analysis is carried out within the Furry picture to first order in the fine-structure constant. It is shown that the presence of photons in the initial state gives rise to an additional (stimulated) channel of photon emission besides the pure vacuum one. On the other hand, the number of final (signal) photons includes also a negative contribution due to photon absorption within the pair production process. These contributions are evaluated and compared. To obtain quantitative predictions in the domain of realistic field parameters, we employ the WKB approach. We propose to use an optical probe photon beam, whose intensity changes as it traverses a spatial region where a strong electric component of a background laser field is present. It is demonstrated that relative intensity changes on the level of $1 \%$ can be experimentally observed once the intensity of the strong background field exceeds $10^{27}~\text{W/cm}^2$ within a large laser wavelength interval. This finding is expected to significantly support possible experimental investigations of nonlinear QED phenomena in the nonperturbative regime.
\end{abstract}

\maketitle

\section{Introduction}\label{sec:intro}

As early as the 1930s~\cite{euler_kockel, heisenberg_euler, weisskopf}, it became evident that the quantum nature of the electromagnetic interaction manifests itself in an effective violation of the superposition principle taking place in the classical theory based on Maxwell's equations in vacuum. It was found that Maxwell's Lagrangian gains additional quantum corrections which lead to remarkable nonlinear phenomena such as light-by-light scattering~\cite{euler_kockel, weisskopf, heisenberg_euler, karplus_pr_1950_1951} and Sauter-Schwinger electron-positron pair production~\cite{sauter_1931, heisenberg_euler, schwinger_1951} (for review, see, e.g., Refs.~\cite{dittrich_gies, dunne_shifman, dunne_epjd_2009, heinzl_ilderton_epjd_2009, marklund_epjd_2009, dipiazza_rmp_2012, king_heinzl_2016, xie_review_2017, blaschke_review_2016}). To galvanize the quantum fluctuations, one basically strives to make them interact with a strong background field. In this context, the rapid developments of the technology for generating high-power laser pulses has continuously encouraged active theoretical and experimental research. Although some of the nonlinear phenomena of strong-field quantum electrodynamics (QED) were already practically observed~\cite{burke_prl_1997, bamber_prd_1999, cole_prx_2018, poder_prx_2018}, the nonperturbative process of electron-positron pair production, i.e., the Sauter-Schwinger effect, is still experimentally unexplored. Whereas one can investigate analogous phenomena in condensed matter systems (see, e.g., Refs.~\cite{szpak_njp_2012, fillion_prb_2015, akal_prd_2016, kasper_njp_2017, linder_prb_2018, pineiro_njp_2019, smolyansky_particles_2020, solinas_prl_2021}), in standard QED the Sauter-Schwinger mechanism is exponentially suppressed unless the electric field strength approaches the critical value $E_\text{c} = m^2 c^3 /(|e|\hbar) \approx 1.3 \times 10^{16}~\text{V/cm}$ ($m$ and $e<0$ are the electron mass and charge, respectively). This corresponds to an intensity of $2.3 \times 10^{29}~\text{W/cm}^2$, while the maximum intensity achieved so far amounts to $10^{23}~\text{W/cm}^2$~\cite{yoon_2021}.

\begin{figure}[b]
\center{\includegraphics[width=0.5\linewidth]{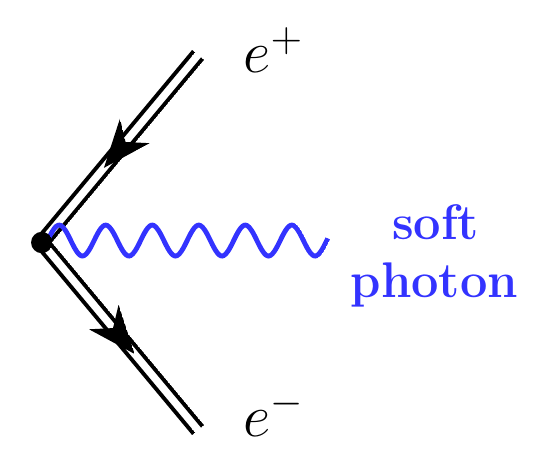}}
\caption{Vertex diagram describing vacuum emission of soft photons accompanying the Sauter-Schwinger effect. The double lines represent the exact electron wavefunction in the presence of the external field, i.e., the external classical background is treated nonperturbatively.}
\label{fig:vertex}
\end{figure}

As was proposed in Ref.~\cite{otto_prd_2017}, one can attempt to experimentally study the process of vacuum photon emission accompanying the Sauter-Schwinger mechanism as illustrated in Fig.~\ref{fig:vertex}. Measuring this additional radiation would allow one to indirectly probe nonperturbative pair production. This vertex diagram exactly incorporates the interaction with the classical electromagnetic background, which is reflected by the double fermionic lines. The process described by the diagram is of the first order in the fine-structure constant $\alpha = e^2/(4\pi \hbar c)$ and predicts emission of a huge number of soft photons~\cite{di_piazza_prd_2005, otto_prd_2017, aleksandrov_prd_2019_2}. In the present study, we revisit the quantitative features of this phenomenon providing closed-form expressions and numerical estimates describing the number density of photons emitted in the domain of realistic field parameters. Furthermore, we propose another experimental scenario which is closely related to vacuum photon emission but involves additional (probe) photons already in the initial state. It turns out that the presence of these photons induces two additional contributions besides the purely vacuum one discussed above. One of them is exactly the same diagram as that displayed in Fig.~\ref{fig:vertex} with the outgoing-photon state coinciding with that of the initial (probe) photon. The other contribution is negative and describes the absorption of the initial photon with the production of an electron-positron pair. As will be shown in what follows, the sum of these two photon-induced terms can lead to notable change of the probe beam intensity. Measuring this change is another tool for investigating the Sauter-Schwinger mechanism in the experiment. In this paper, we argue that this scenario proves to be more favorable than measuring the vacuum radiation itself or detecting pairs directly since the probe-beam technique corresponds to a lower threshold with respect to the laser intensity.

Since we are interested in the nonperturbative regime, the interaction with the classical external field is taken into account exactly, i.e., we work within the Furry picture. The quantized part of the electromagnetic field is treated within perturbation theory (PT). To compute the Feynman diagrams, we first employ our numerical technique developed previously in the context of pair production~\cite{aleksandrov_prd_2016, aleksandrov_prd_2017_2, aleksandrov_prd_2018} and subsequently generalized for studying radiative processes~\cite{aleksandrov_prd_2019_2}. Second, we also perform calculations using perturbation theory with respect to the classical background in order to benchmark our nonperturbative numerical procedures. Finally, to obtain quantitative predictions in the domain of realistic field parameters, we construct the necessary wavefunctions by means of the WKB approach and complete the evaluation of the diagrams analytically. The closed-form final expressions are then used to examine the experimental feasibility of our proposal.

The paper has the following structure. In Sec.~\ref{sec:gen} we recap the main general features of vacuum photon emission. In Sec.~\ref{sec:photons} we discuss the photon-induced contributions in the case of one initial photon or many identical photons. Moving on to specific calculations, in Sec.~\ref{sec:calc_pt} we first describe our nonperturbative procedure in more detail and then employ PT to benchmark our technique. In Sec.~\ref{sec:scalar} we examine the photon emission process within the framework of scalar QED. In Sec.~\ref{sec:wkb} we perform WKB calculations and obtain closed-form expressions for the necessary quantities. The experimental prospects of our proposal as well as the feasibility of measuring the vacuum radiation and Sauter-Schwinger mechanism itself are discussed in Sec.~\ref{sec:exp}. Finally, we conclude in Sec.~\ref{sec:conclusions}. We will employ the units $\hbar = c = 1$.

\section{Vacuum photon emission within the Furry picture} \label{sec:gen}

In our study, the external classical field is treated nonperturbatively, i.e., within the Furry picture. The quantized electron-positron field $\psi$ interacts with both the classical background $\mathcal{A}^\mu$ and quantized part of the electromagnetic field $\hat{A}^\mu$. The photons emitted as well as the probe photons are quanta of the latter. The quantized part is incorporated by PT within the interaction picture~\cite{fradkin_gitman_shvartsman}. The corresponding $S$ operator has the form
\begin{equation}
S = T~\mathrm{exp} \Bigg ( \!\! -i \int d^4 x \ j^\mu (x) \hat{A}_\mu (x) \Bigg ), \label{eq:S_def}
\end{equation}
where $x = (t, \boldsymbol{x})$, $T$ is the time-ordering operator, and $j^\mu$ is a current operator in the presence of the external background $\mathcal{A}^\mu$. The external field strength is assumed to vanish outside the interval $t \in [t_\text{in},\ t_\text{out}]$, and in Eq.~\eqref{eq:S_def} one integrates over this temporal region. In our case the field is switched on and off adiabatically, so we will imply $t_\text{in/out} \to \mp \infty$.

The quantized part of the electromagnetic field has the following standard decomposition in terms of the photon mode functions:
\begin{equation}
\hat{A}_\mu (x) = \sum_{\lambda=0}^3 \int \! d\boldsymbol{k} \, \Big [ c_{\boldsymbol{k}, \lambda} f_{\boldsymbol{k},\lambda, \mu} (x) + c^\dagger_{\boldsymbol{k}, \lambda} f^*_{\boldsymbol{k},\lambda, \mu} (x) \Big ],
\label{eq:photon_field}
\end{equation}
where $c^\dagger_{\boldsymbol{k}, \lambda}$ and $c_{\boldsymbol{k}, \lambda}$ are the photon creation and annihilation operators, respectively, and $f_{\boldsymbol{k},\lambda,\mu} (x) = (2\pi)^{-3/2} (2k^0)^{-1/2} \, \mathrm{e}^{-ikx} \varepsilon_\mu (\boldsymbol{k}, \lambda)$ is the photon wavefunction corresponding to momentum $\boldsymbol{k}$ ($k^0 = |\boldsymbol{k}|$) and polarization $\lambda$. The electron-positron field operator $\psi$ can be decomposed either in terms of the so-called {\it in} one-particle solutions ${}_\pm \varphi_n (x)$ or in terms of the {\it out} solutions ${}^\pm \varphi_n (x)$. The {\it in} ({\it out}) wavefunctions are determined by their asymptotic form for $t \leqslant t_\text{in}$ ($t \geqslant t_\text{out}$), where they have a well-defined sign of energy denoted by $\pm$. Quantum number $n$ incorporates momentum and spin. In what follows, we will need the expansion of the electron-positron field operator in terms of the {\it in} solutions of the Dirac equation,
\begin{equation}
\psi (x) = \sum_{n} \big [a_n \, {}_+ \varphi_n (x) + b^\dagger_n \, {}_- \varphi_n (x)\big ], \label{eq:psi_x_in}
\end{equation}
where we have introduced the electron (positron) creation and annihilation operators $a^\dagger_n$ ($b^\dagger_n$) and $a_n$ ($b_n$), respectively. These operators obey the usual anticommutation relations. The vacuum state will be denoted by $|0,\text{in}\rangle$. The current operator involved in Eq.~\eqref{eq:S_def} can now be constructed via $j^\mu (x) = (e/2)[\bar{\psi}(x)\gamma^\mu, \, \psi(x)]$.

In this section, we assume that the initial state is a pure vacuum one, i.e., we do not introduce any additional photons, $|\text{in}\rangle = |0,\text{in}\rangle$. To describe the process of photon emission, we will evaluate the number density of photons in the final state, which can be obtained by evolving $|0,\text{in}\rangle$ with the aid of the $S$ operator~\eqref{eq:S_def}:
\begin{equation}
n_{\boldsymbol{k}, \lambda}^{\text{(vac)}} = \langle 0,\text{in} | S^\dagger c^\dagger_{\boldsymbol{k}, \lambda} c_{\boldsymbol{k}, \lambda} S |0,\text{in}\rangle.
\label{eq:ph_number_density_int}
\end{equation}
Note that this quantity is an inclusive observable, i.e., the final state is not specified here and it may contain an arbitrary number of pairs. The ``mean number'' $n_{\boldsymbol{k}, \lambda}^{\text{(vac)}}$ represents the number density of photons in the momentum space, $n_{\boldsymbol{k}, \lambda}^{\text{(vac)}} = dN_{\boldsymbol{k}, \lambda}^{\text{(vac)}}/d\boldsymbol{k}$. In the present study, we evaluate the expression~\eqref{eq:ph_number_density_int} to first order in the fine-structure constant using the series expansion of the exponential~\eqref{eq:S_def}. Straightforward calculations yield~\cite{aleksandrov_prd_2019_2}:
\begin{eqnarray}
n_{\boldsymbol{k}, \lambda}^{\text{(vac)}} &=& \bigg | \int \! d^4x \, j_{\text{in}}^\mu (x) f^*_{\boldsymbol{k},\lambda,\mu} (x) \bigg |^2 \nonumber \\
{} &+& e^2 \sum_{n,m} \bigg | \int \! d^4x \, {}_+ \bar{\varphi}_n (x) \gamma^\mu f^*_{\boldsymbol{k},\lambda,\mu} (x) \, {}_- \varphi_m (x) \bigg |^2, \label{eq:ph_number_density_gen_sum}
\end{eqnarray}
where $j_{\text{in}}^\mu (x) = \langle 0,\text{in} | j^\mu (x) |0,\text{in}\rangle$ is the vacuum current. This expression can also be derived by computing transition amplitudes and summing over the possible final states containing various numbers of electrons and positrons. This alternative, if tedious, scheme is presented in Ref.~\cite{fradkin_gitman_shvartsman}.

The first term in Eq.~\eqref{eq:ph_number_density_gen_sum} corresponds to the so-called tadpole (reducible) contribution explored in numerous studies (see, e.g., Refs.~\cite{di_piazza_prd_2005, aleksandrov_prd_2019_2, fradkin_gitman_shvartsman, fedotov_pla_2006, fedotov_las_2007, karbstein_prd_2015, gies_prd_2018_1, gies_prd_2018_2, king_pra_2018, blinne_prd_2019, karbstein_prl_2019_2, aleksandrov_pra_2021}). It predicts emission of photons similar to those constituting the external field or higher harmonics. The most robust technique for computing this term is based on the locally-constant field approximation (LCFA)~\cite{karbstein_prd_2015, gies_prd_2018_1, gies_prd_2018_2, king_pra_2018, blinne_prd_2019, karbstein_prl_2019_2}. This technique is expected to be accurate once the external-field frequency $\omega$ is much less than $m$, which was evidently confirmed in our recent investigation~\cite{aleksandrov_prd_2019_2}, where it was also shown that the LCFA prediction may considerably differ from the exact values of the photon yield if $(\omega/m)^2 \gtrsim 0.3$.

In this study, we focus on the process of soft photon emission accompanying the Sauter-Schwinger mechanism of pair production. This process is described by the second term in Eq.~\eqref{eq:ph_number_density_gen_sum}, which can be illustrated by the vertex diagram in Fig.~\ref{fig:vertex}. Since it is responsible for the low-energy part of the radiation spectrum, we refrain from discussing the tadpole contributions in what follows. In Ref.~\cite{otto_prd_2017} the vertex diagram was examined in the case of a spatially uniform external background. Recently~\cite{aleksandrov_prd_2019_2}, it was demonstrated that taking into account the spatiotemporal inhomogeneities of the external field in the case of a standing electromagnetic wave leads to a notable anisotropy of the emitted photons providing additional signatures that can be, in principle, measured in the experiment.

Here we will also assume that the external field does not depend on the spatial coordinates, $\mathcal{A}^\mu = \mathcal{A}^\mu (t)$. This will allow us to obtain relatively simple estimates approximating a combination of two counterpropagating (high-intensity) laser pulses in the vicinity of a maximal electric field amplitude by a uniform background. As was shown in Refs.~\cite{otto_prd_2017, aleksandrov_prd_2019_2}, the photon number density for low energies $k_0 \ll m$ is proportional to $1/k_0^3$. In the case of a spatially homogeneous field, Eq.~\eqref{eq:ph_number_density_gen_sum} yields factor $V$, the volume of the system, so we isolate it and present the photon number density in the following form:
\begin{equation}
\frac{n_{\boldsymbol{k}, \lambda}^{\text{(vac)}}}{V} = \frac{A_{\boldsymbol{n},\lambda}}{k_0^3} + \frac{B_{\boldsymbol{n},\lambda}}{k_0^2} + ..., \label{eq:A_B}
\end{equation}
where $\boldsymbol{n} = \boldsymbol{k}/k_0$, i.e., the coefficients $A_{\boldsymbol{n},\lambda}$ and $B_{\boldsymbol{n},\lambda}$ depend on the photon polarization $\lambda$ and the propagation direction. Note that although the photon number density diverges as $k_0 \to 0$, the energy emitted is finite. The function $f_\gamma (\boldsymbol{k})$ introduced in Ref.~\cite{otto_prd_2017} corresponds to the sum of Eq.~\eqref{eq:A_B} over $\lambda$. In quantitative estimates, the interaction volume $V$ is treated as a volume of the focal spot of laser radiation. The experimental prospects will be discussed in detail in Sec.~\ref{sec:exp}.

One might also ask whether it is necessary to take into account the vertex diagram in Fig.~\ref{fig:vertex} when computing the number of electron-positron pairs. In this case, one has to integrate over the photon momentum $\boldsymbol{k}$, which leads to an infrared divergence. However, according to Ref.~\cite{bloch_nordsieck_1937}, the divergent higher-order contributions, in fact, do not affect the leading order result (see also Ref.~\cite{blp}; this issue was examined in the context of other nonperturbative processes in, e.g., Refs.~\cite{seipt_kaempfer_prd_2012, ilderton_prd_2013, ilderton_plb_2013}). The same holds true with regard to radiative corrections to the process of soft photon emission considered in the present study. As we always specify the signal photon energy $k_0 > 0$, the leading contribution~\eqref{eq:A_B} is finite, whereas the higher-order infrared divergences are irrelevant. On the other hand, the number of photons can be affected by higher-order diagrams where multiple $k_0$-quanta are emitted simultaneously. Nevertheless, these corrections are small since they are suppressed by powers of $\alpha \ln (m/k_0)$~\cite{bloch_nordsieck_1937, blp}. In what follows, we will assume that $k_0 \geqslant 10^{-6} m$, which leads to $\alpha \ln (m/k_0) < 0.1$, so the higher-order terms will be disregarded.

\section{Additional photons in the initial state} \label{sec:photons}

In this section, we will analyze the process of photon emission in the presence of additional (probe) photons in the initial state, which were absent in the vacuum state $|\text{0, in} \rangle$ considered previously.

\begin{figure}[b]
\center{\includegraphics[width=0.95\linewidth]{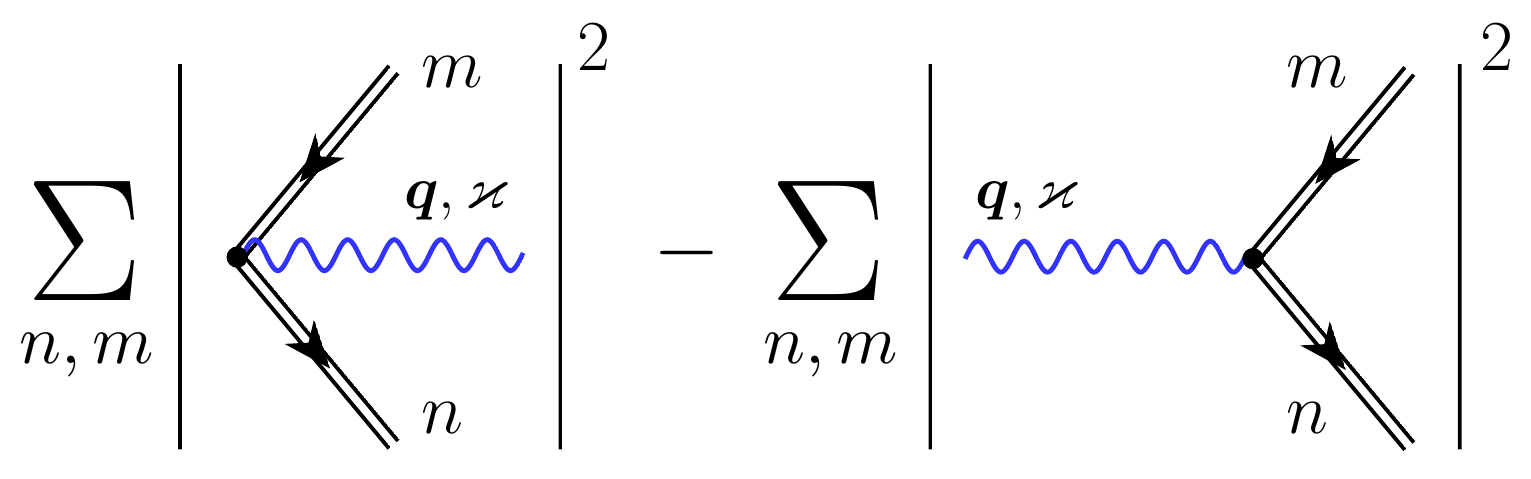}}
\caption{Two additional contributions which appear in the case of a one-photon initial state. The sum over the final fermionic states involves integration over momentum and summation over spin. The photon lines correspond to the quantum numbers $\boldsymbol{q}$ and $\varkappa$ of the initial photon.}
\label{fig:diff}
\end{figure}

\subsection{One-photon state} \label{sec:photons_one}

Consider now a one-photon state $|\text{in} \rangle = c^\dagger_{\boldsymbol{q}, \varkappa} |\text{0, in} \rangle$. Calculating the photon number density to first order in $\alpha$ via 
\begin{equation}
n^{(1)}_{\boldsymbol{k}, \lambda} = \langle 0,\text{in} | c_{\boldsymbol{q}, \varkappa} S^\dagger c^\dagger_{\boldsymbol{k}, \lambda} c_{\boldsymbol{k}, \lambda} S c^\dagger_{\boldsymbol{q}, \varkappa} |0,\text{in}\rangle,
\label{eq:ph_number_density_induced}
\end{equation}
one obtains a sum of two contributions. The first one is exactly $n_{\boldsymbol{k}, \lambda}^{\text{(vac)}}$ given in Eq.~\eqref{eq:ph_number_density_gen_sum}. Note that this part does not involve the quantum numbers of the probe photon. The second contribution to $n_{\boldsymbol{k}, \lambda}$ appears due to the presence of the additional photon and represents the difference shown in Fig.~\ref{fig:diff}. The photon wavefunctions correspond to the quantum numbers $\boldsymbol{q}$ and $\varkappa$ of the initial photon, so these terms contribute only when $\boldsymbol{k}$, $\lambda$ coincide with $\boldsymbol{q}$, $\varkappa$. Together with the process without additional quanta, the three parts can be interpreted as spontaneous emission, stimulated emission, and photon absorption, respectively, similarly to the atomic physics notions. To properly treat the ``resonance'' character of the photon-induced terms, one should introduce a smearing function for the initial photon state. We assume that it is localized in a small vicinity of $\boldsymbol{q}$ which has volume $V_q$ (in momentum space). Another important point is the fact that the difference between the two diagrams in Fig.~\ref{fig:diff} appears due to complex conjugation of the photon wavefunction, which is equivalent to the substitution $k^\mu \to -k^\mu$. It means that the second diagram in Fig.~\ref{fig:diff} (photon absorption) has the same behavior~\eqref{eq:A_B} with the opposite sign of the $1/k_0^2$ term. One factor $1/k_0$ relates to the normalization of the photon wavefunctions, so one should change the sign of the even powers. Whereas the difference in Fig.~\ref{fig:diff} no longer has the $1/k_0^3$ term, the next-to-leading-order term doubles. Note that there are no stimulated tadpole contributions, which is no surprise as the vacuum current is real, so conjugating the photon wavefunction, one does not change the absolute value of the diagram.

We assume that the probe photons will be measured within the whole $V_q$ region, i.e., we integrate the photon number density over the small momentum volume, where the initial photon was localized. It brings us to the following number of photons ($\boldsymbol{l} = \boldsymbol{q}/q_0$):
\begin{equation}
N^{(1)} = 1 + \bigg ( \frac{A_{\boldsymbol{l},\varkappa}}{q_0^3} + \frac{B_{\boldsymbol{l},\varkappa}}{q_0^2} + ... \bigg ) \, V V_q + \frac{2 (2\pi)^3 B_{\boldsymbol{l},\varkappa}}{q_0^2} + ... \label{eq:N_one_photon}
\end{equation}
Here the first term corresponds to the trivial zeroth-order contribution and merely indicates that the initial state contains one photon. The second vacuum term is enhanced by the large factor $V$ since the vacuum emission takes part in the whole interaction region. On the other hand, it is suppressed by $V_q$ since we have integrated only over this small momentum region. In the quantitative estimates concerning this vacuum term, we will take into account that this radiation is emitted at all the other directions (see Sec.~\ref{sec:exp}). The vacuum contribution is essentially determined by the first term in parentheses in Eq.~\eqref{eq:N_one_photon}. Finally, the last term in Eq.~\eqref{eq:N_one_photon} comes from the difference in Fig.~\ref{fig:diff} and governs the photon-induced contribution. Here the factor $2$ appears since the $B$ term doubles in the difference of the two Feynman diagrams. Let us briefly discuss the origin of the factor $(2\pi)^3$. When introducing a photon smearing function, one has to perform two additional integrations in Eq.~\eqref{eq:ph_number_density_induced}. Assuming that the smearing function is equal to some constant $C$ in a small vicinity of momentum $\boldsymbol{q}$, one can simply multiply the integrand by factor $V_q^2$ and replace the smearing function with $C$. This leads to an overall factor $C^2V_q^2$. Since the initial state contains one quantum, the zeroth-order photon number density integrated over $V_q$ yields unity, i.e, $C^2 V_q = 1$. The external field is homogeneous in space, so the number density will be proportional to the volume $V$ as in Eq.~\eqref{eq:A_B}. However, here $V$ is determined by the spatial volume occupied by the initial photon, i.e., $V$ should be replaced with $(2\pi)^3/V_q$. Accordingly, we arrive at the factor $(2\pi)^3$ in the last term in Eq.~\eqref{eq:N_one_photon}. Note that unlike the vacuum contribution, this one is enhanced once the initial state contains many probe photons. This will be discussed next.

\subsection{$N$-photon state} \label{sec:photons_N}

The stimulated emission and absorption parts give rise to the number of signal photons corresponding to the third term in Eq.~\eqref{eq:N_one_photon}. It is not enhanced by $V$ since the probe photon does not ``feel'' the boundaries of the interaction region. However, the crucial point is that in the case of $N$ photons in the initial state, this term will be proportional to $N$, which can be verified by direct calculations of the mean number density of photons as was discussed above. Accordingly, the photon-induced contribution yields
\begin{equation}
N^{\text{(ph)}} = \frac{2 (2\pi)^3 |B_{\boldsymbol{l},\varkappa}|}{q_0^2} \, N.
\label{eq:N_ph_final}
\end{equation}
As will be seen later, the coefficient $B_{\boldsymbol{l},\varkappa}$ can be positive or negative depending on the field configuration, i.e., the dominant contribution can arise from either the process of stimulated emission or from the absorption channel.

Here we propose to measure the relative change of the optical probe beam intensity given by the ratio $N^{\text{(ph)}}/N$. In Sec.~\ref{sec:exp} we will assess this scenario together with the proposal to detect the vacuum radiation and with possible observations of the pair production process. In order to provide quantitative estimates and discuss the feasibility of measuring the signal experimentally, one should evaluate the coefficients $A_{\boldsymbol{n}, \lambda}$ and $B_{\boldsymbol{n}, \lambda}$. This will be a subject of the following sections.

\section{Furry-picture calculations versus perturbation theory} \label{sec:calc_pt}

In this section, we will perform numerical calculations of the coefficients $A_{\boldsymbol{n}, \lambda}$ and $B_{\boldsymbol{n}, \lambda}$ and benchmark the results by means of PT. In what follows, the external background is assumed to be a Sauter pulse
\begin{equation}
E_z(t) = \frac{E_0}{\cosh^2 (t/\tau)},
\label{eq:ext_field_sauter}
\end{equation}
so the classical potential reads
\begin{equation}
\mathcal{A}^3 (t) = -E_0 \tau \tanh (t/\tau).
\label{eq:ext_field_potential}
\end{equation}
The other components vanish. We also introduce $\mathcal{A}_0 \equiv \mathcal{A}^3 (+\infty) = -E_0 \tau$ and $\boldsymbol{\mathcal{A}}_0 = \mathcal{A}_0 \boldsymbol{e}_z$.

\subsection{Nonperturbative calculations} \label{sec:calc_pt_calc}

To calculate the coefficients $A_{\boldsymbol{n}, \lambda}$ and $B_{\boldsymbol{n}, \lambda}$, one has to compute the second term in Eq.~\eqref{eq:ph_number_density_gen_sum}. Since the external field does not depend on the spatial coordinates, the in solutions involved in Eq.~\eqref{eq:ph_number_density_gen_sum} can be represented as
\begin{equation}
{}_\zeta \varphi_{\boldsymbol{p},s} (x) = (2\pi)^{-3/2} \, \mathrm{e}^{i \zeta \boldsymbol{p} \boldsymbol{x}} {}_\zeta \chi_{\boldsymbol{p},s} (t),
\label{eq:solutions_chi}
\end{equation}
where $\zeta = \pm$ and $s = \pm 1$ defines the spin state. Then the vacuum contribution to the number density of soft photons takes the following form~\cite{otto_prd_2017, aleksandrov_prd_2019_2}:
\begin{widetext}
\begin{equation}
\frac{(2\pi)^3}{V} \, n_{\boldsymbol{k}, \lambda}^{\text{(vac)}} = \frac{\alpha}{4\pi^2} \frac{1}{k_0} \sum_{s, s'} \int \! d\boldsymbol{p} \bigg | \int \! dt \, {}_+ \bar{\chi}_{\boldsymbol{p},s} (t) \gamma^\mu \varepsilon^*_\mu (\boldsymbol{k},\lambda) \, {}_- \chi_{-\boldsymbol{p}-\boldsymbol{k}, s'} (t) \mathrm{e}^{ik_0 t} \bigg |^2,
\label{eq:ph_number_density_uniform}
\end{equation}
\end{widetext}
When integrating over $t \in (-\infty, t_\text{in}]$ and $t \in [t_\text{out}, +\infty)$, one has to introduce a factor $\mathrm{e}^{-\varepsilon |t|}$ ($\varepsilon \to 0$). In fact, the leading contribution $A_{\boldsymbol{n}, \lambda}/k_0^3$ arises from the region $[t_\text{out}, +\infty)$~\cite{otto_prd_2017}, so one does not need to perform numerical integration over the intermediate time domain. Moreover, the coefficient $A_{\boldsymbol{n}, \lambda}$ can be evaluated analytically (see Appendix~A),
\begin{equation}
A_{\boldsymbol{n}, \lambda} = \frac{\alpha}{4\pi^5} \int \! d\boldsymbol{p} \, \frac{(\boldsymbol{P},\boldsymbol{e}_\lambda)^2 p_0^2 (\boldsymbol{P})}{\big [p_0^2 (\boldsymbol{P}) - (\boldsymbol{P},\boldsymbol{n})^2 \big ]^2} \, n_{\boldsymbol{p}} (1 - n_{\boldsymbol{p}}),
\label{eq:A_spinor_exact}
\end{equation}
where $\boldsymbol{P} \equiv \boldsymbol{p} - e \boldsymbol{\mathcal{A}}_0$, $p_0 (\boldsymbol{p}) \equiv \sqrt{m^2 + \boldsymbol{p}^2}$, and $\boldsymbol{e}_\lambda$ is a three-dimensional photon polarization vector. To derive the expression~\eqref{eq:A_spinor_exact}, we have employed the explicit formulas for the one-particle transitions, which can be obtained in the case of the Sauter pulse~\eqref{eq:ext_field_sauter} since the Dirac equation can be solved analytically~\cite{fradkin_gitman_shvartsman, narozhny_1970}. The number density $n_{\boldsymbol{p}}$ of the electrons produced is also known exactly~\cite{fradkin_gitman_shvartsman, narozhny_1970, gavrilov_prd_1996}. Note that this quantity does not depend on spin $s$ and never esceeds unity.

To evaluate the coefficient $B_{\boldsymbol{n}, \lambda}$, we use the general expression~\eqref{eq:ph_number_density_uniform} and subtract the term with the opposite sign of $k^\mu$ from the $\boldsymbol{p}$ integrand and then divide the result by~2.

\subsection{Perturbation theory} \label{sec:calc_pt_pt}

\begin{figure}[t]
\center{\includegraphics[width=0.99\linewidth]{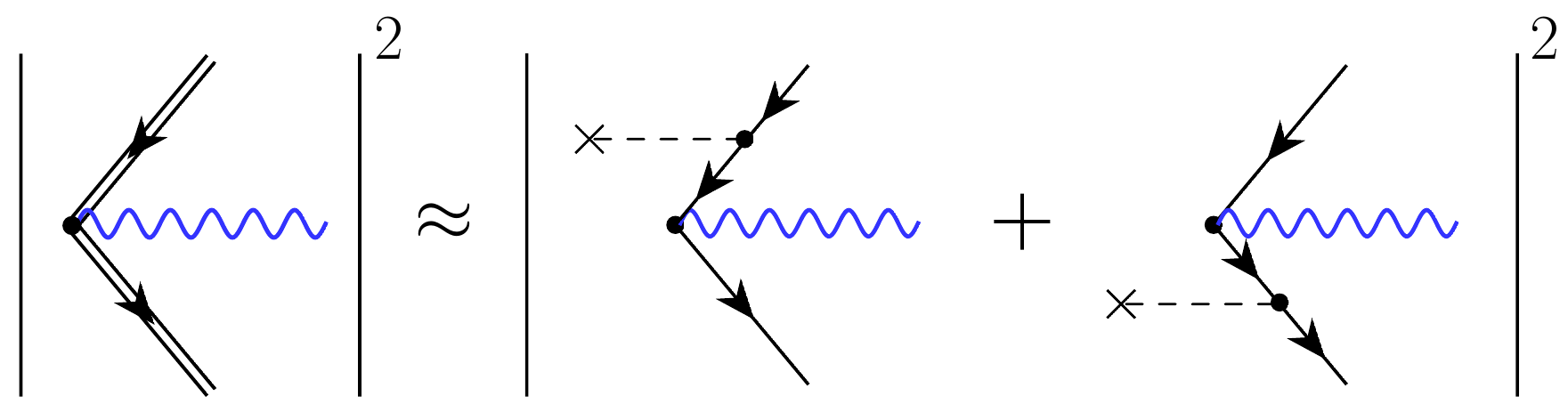}}
\caption{Perturbative expansion of the vertex diagram.}
\label{fig:pt}
\end{figure}

We also employ a PT approach to test our nonperturbative procedure. The leading contribution of the vertex diagram can be evaluated as displayed in Fig.~\ref{fig:pt}. The ordinary thin lines correspond here to the free solutions of the Dirac equation or to the free propagator in the case of the internal line. The crosses denote the interaction with the classical background $\mathcal{A}^\mu$.

Computing directly the Feynman diagrams in Fig.~\ref{fig:pt}, we arrive at
\begin{equation}
A^{(\text{PT})}_{\boldsymbol{n}, \lambda} = \frac{\alpha}{4\pi^3} (eE_0 \tau^2)^2 \int \! d\boldsymbol{p} \, \frac{1}{\sinh^2 (\pi \tau p_0)} \, \frac{(\boldsymbol{p},\boldsymbol{e}_\lambda)^2 (p_0^2 - p_z^2)}{\big [p_0^2 - (\boldsymbol{p},\boldsymbol{n})^2 \big ]^2},
\label{eq:A_spinor_pt}
\end{equation}
where $p_0 = p_0(\boldsymbol{p}) = \sqrt{m^2 + \boldsymbol{p}^2}$. This expression coincides with the leading term of the perturbative expansion of Eq.~\eqref{eq:A_spinor_exact}, as it should. To calculate $B_{\boldsymbol{n}, \lambda}$, we utilize the subtraction procedure as described in the previous section. The PT approach is only accurate if $|eE_0|\tau \ll m$, i.e., the so-called Keldysh parameter $\gamma = m/(|eE_0|\tau)$ is sufficiently large, which is obviously not a realistic condition. For this reason, one has to perform nonperturbative calculations as was stated above.

\begin{figure*}[t]
\center{\includegraphics[height=0.35\linewidth]{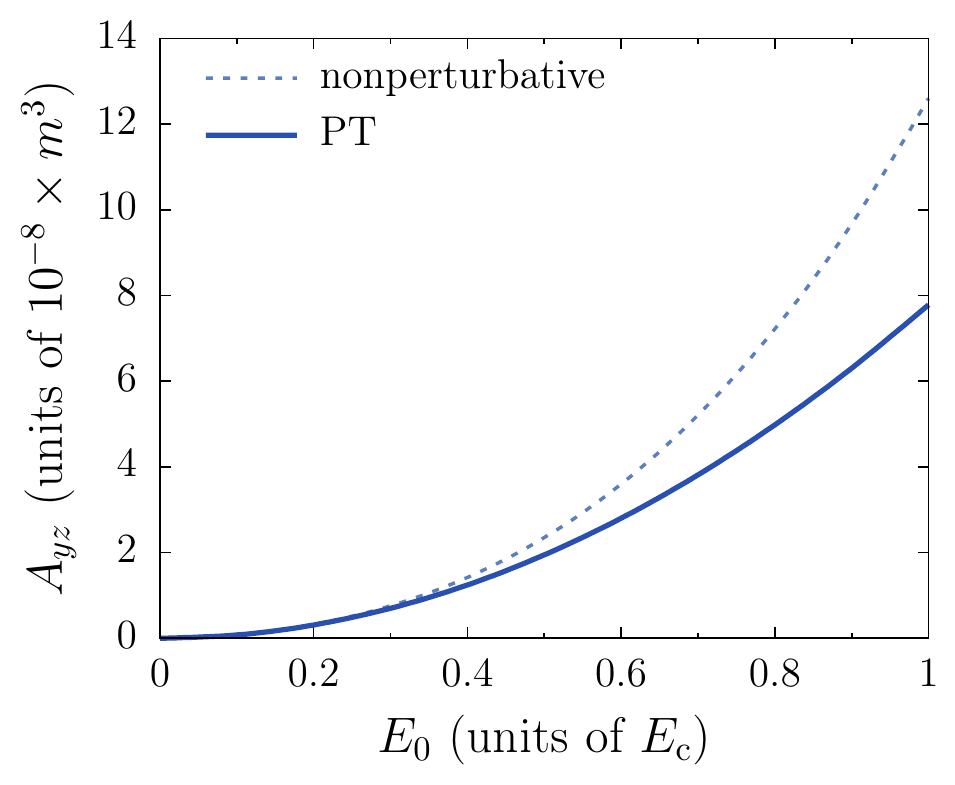}~~~~~~~~~~~~\includegraphics[height=0.35\linewidth]{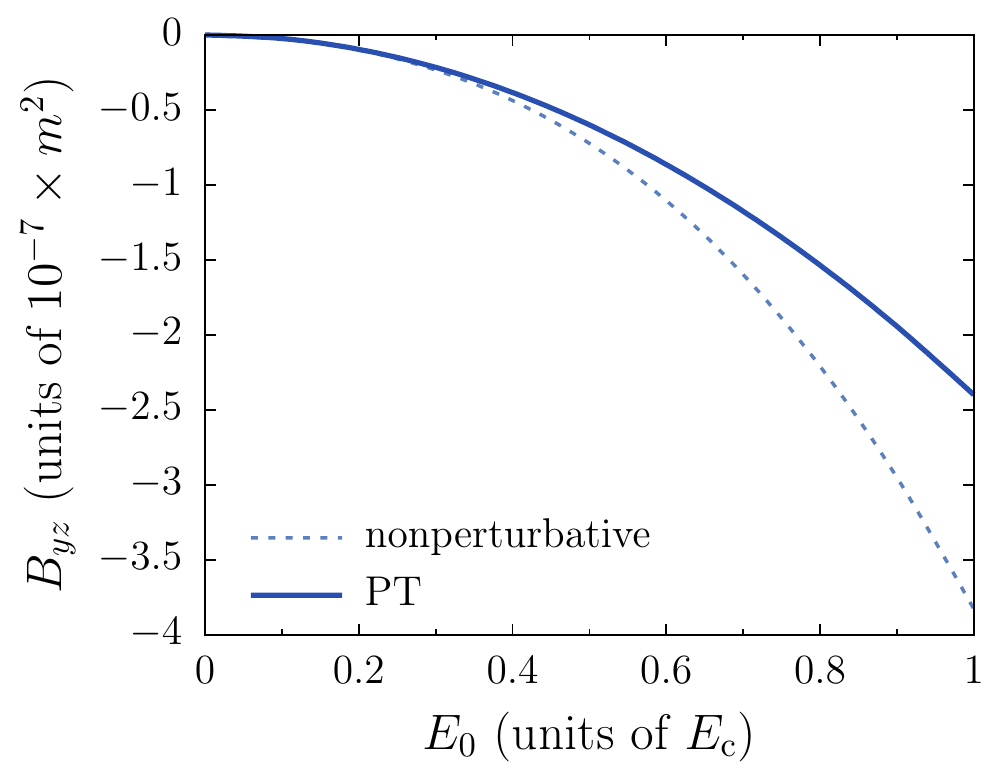}}
\caption{Coefficients $A_{yz}$ and $B_{yz}$ as functions of the field amplitude $E_0$ of the Sauter pulse~\eqref{eq:ext_field_sauter} for $\tau = 1.0 m^{-1}$. The results were obtained by means of the nonperturbative expression~\eqref{eq:ph_number_density_uniform} and within perturbation theory to second order in $E_0$.}
\label{fig:res_pt}
\end{figure*}

However, to benchmark our exact numerical approach, we compare the results with the PT predictions. As an example, in Fig.~\ref{fig:res_pt} we display the coefficients $A_{\boldsymbol{n}, \lambda}$ and $B_{\boldsymbol{n}, \lambda}$ for $\boldsymbol{n} = \boldsymbol{e}_y$ and photon polarization along the $z$ axis, $\boldsymbol{e}_\lambda = \boldsymbol{e}_z$ (we use index $yz$). The coefficients are presented as functions of $E_0$ for $\tau = 1.0 m^{-1}$. We observe that the PT approach provides quite accurate predictions for sufficiently weak pulses.

Finally, we point out that unlike the PT approach, one cannot employ the so-called locally-constant field approximation (LCFA) as it is not suitable for computing the number density of soft photons. In fact, one usually utilizes the LCFA, for instance, to describe the nonlinear Breit-Wheeler mechanism of high-energy-photon decay in strong external fields. Note that according to Refs.~\cite{reiss_1962, nikishov_ritus_jetp_1964}, in the case of a plane wave background the second diagram in Fig.~\ref{fig:diff} is exponentially suppressed. The first diagram in Fig.~\ref{fig:diff} does not even contribute in a plane wave in accordance with the fact that plane waves do not produce pairs. Our direct nonperturbative calculations taking into account the temporal dependence of the external field capture the effect of interest unlike the LCFA. To incorporate spatial inhomogeneities, one should either evaluate diagrams in the presence of space-time-dependent fields, which is a formidable task, or sum the results over the spatial profile according to the local approximation employed in Ref.~\cite{aleksandrov_prd_2019_2} (see also Ref.~\cite{aleksandrov_kohlfuerst_prd_2020}). However, here we assume the field to be spatially uniform to save computational time and obtain the necessary estimates. The spatial finiteness of the interaction region is taken into account by the volume factor $V$ as will be discussed in Sec.~\ref{sec:exp}.

\section{Photon emission in scalar QED} \label{sec:scalar}

Since we are interested in computing the coefficients $A_{\boldsymbol{n},\lambda}$ and $B_{\boldsymbol{n},\lambda}$ for more realistic parameters of the external field, it is possible to simplify calculations by using the WKB approach. This is particularly important for calculating $B_{\boldsymbol{n},\lambda}$ as in this case, we do not have a closed-form expression like Eq.~\eqref{eq:A_spinor_exact}. As we assume that the external field is linearly polarized and spatially homogeneous, the spin effects should be insignificant, which allows one to consider scalar QED, where the calculations are simpler. The result will be multiplied by a factor of 2. Note that the total particle yield in spinor and scalar QED can possess different quantitative patterns also due to effects of statistics~\cite{sevostyanov_prd_2021}. However, these effects come into play only for $E_0 \gtrsim E_\text{c}$, so here they can be completely disregarded.

It turns out that in scalar QED the general expression for the vertex contribution is similar to the second term in Eq.~\eqref{eq:ph_number_density_gen_sum}. It reads
\begin{equation}
n_{\boldsymbol{k}, \lambda}^{\text{(sc)}} = e^2 \sum_{n,m} \bigg | \int \! d^4x f^*_{\boldsymbol{k},\lambda,\mu} (x) \, {}_+ \varphi^*_n (x) \overleftrightarrow{\partial^\mu} \, {}_- \varphi_m (x) \bigg |^2,
\label{eq:gen_vac_scalar}
\end{equation}
where $\varphi_1 \overleftrightarrow{\partial^\mu} \varphi_2 \equiv \varphi_1 (\partial^\mu \varphi_2) - (\partial^\mu \varphi_1)\varphi_2$. The quantum numbers $n$ and $m$ correspond to momentum only. In the case of a spatially homogeneous background, the solutions of the Klein-Fock-Gordon equation can be represented as
\begin{equation}
{}_\zeta \varphi_n (x) = {}_\zeta \varphi_{\boldsymbol{p}} (x) = \frac{1}{(2\pi)^{3/2}} \, \mathrm{e}^{\zeta i \boldsymbol{p} \boldsymbol{x}} \, {}_\zeta \chi_{\boldsymbol{p}} (t).
\label{eq:scalar_chi}
\end{equation}
Then Eq.~\eqref{eq:gen_vac_scalar} takes the following form:
\begin{widetext}
\begin{equation}
\frac{(2\pi)^3}{V} \, n_{\boldsymbol{k}, \lambda}^{\text{(sc)}} = \frac{\alpha}{4\pi^2} \frac{1}{k^0} \int \! d\boldsymbol{p} \ \bigg | \int \! dt \, {}_+ \chi^*_{\boldsymbol{p}} (t) \boldsymbol{e}^*_{\lambda} ( 2\boldsymbol{p} + \boldsymbol{k}) {}_- \chi_{-\boldsymbol{p} - {\boldsymbol{k}}} (t) \mathrm{e}^{ik^0 t} \bigg |^2,
\label{eq:scalar_vertex_uniform}
\end{equation}
\end{widetext}
which represents a scalar-QED version of Eq.~\eqref{eq:ph_number_density_uniform}. We assume that $(\boldsymbol{e}_{\lambda}, \boldsymbol{k}) = 0$. The leading $1/k_0^3$ contribution can be evaluated exactly yielding
\begin{equation}
A^{\text{(sc)}}_{\boldsymbol{n}, \lambda} = \frac{\alpha}{8\pi^5} \int \! d\boldsymbol{p} \, \frac{(\boldsymbol{p},\boldsymbol{e}_\lambda)^2 p_0^2 (\boldsymbol{P})}{\big [p_0^2 (\boldsymbol{P}) - (\boldsymbol{P},\boldsymbol{n})^2 \big ]^2} \, n^{\text{(sc)}}_{\boldsymbol{p}} (1 + n^{\text{(sc)}}_{\boldsymbol{p}}),
\label{eq:A_scalar_exact}
\end{equation}
where $n^{\text{(sc)}}_{\boldsymbol{p}}$ is a number density of bosons produced. One observes that in the spinor case, there is also an additional factor of 2 corresponding to the spin degeneracy and the integrand involves $(\boldsymbol{P},\boldsymbol{e}_\lambda)$ instead of $(\boldsymbol{p},\boldsymbol{e}_\lambda)$. The latter point will not be important as we will focus on the case $(\boldsymbol{\mathcal{A}}_0,\boldsymbol{e}_\lambda)=0$. Note that the number density of particles becomes identical in spinor and scalar QED in the realistic regime $E_0 \ll E_\text{c}$, $\tau \gg m^{-1}$, $|eE_0|\tau \gg m$~\cite{gavrilov_prd_1996}. The factor $1 + n^{\text{(sc)}}_{\boldsymbol{p}}$ reflects the statistics of bose particles, cf.~Eq.~\eqref{eq:A_spinor_exact}. In what follows, this factor will be completely inessential as the number density of particles is much smaller than unity in the regime of interest.

Although $B_{\boldsymbol{n},\lambda}$ can be obtained numerically by the subtraction scheme, we will need the results in the realistic domain, where our direct computations become very time consuming. To overcome this obstacle, we will obtain closed-form expressions for $A_{\boldsymbol{n},\lambda}$ and $B_{\boldsymbol{n},\lambda}$ within the WKB approach. Since the spin effects are unimportant here, it is sufficient to carry out the WKB calculations in the case of scalar QED, which is a subject of the next section.

\section{WKB analysis} \label{sec:wkb}

As was stated above, we focus on the realistic domain of the field parameters $E_0 \ll E_\text{c}$, $\tau \gg m^{-1}$, $|eE_0|\tau \gg m$, where the WKB approximation is well justified. As we know the asymptotic behavior of the in solutions for $t \leqslant t_\text{in}$, we can easily calculate the contribution from the region $(-\infty, \, t_\text{in}]$. The wavefunctions for $t>t_\text{in}$ can be constructed approximately. The leading $1/k^3_0$ term comes from the region $[t_\text{out}, \, +\infty)$. To evaluate it, one should decompose the in solutions in Eq.~\eqref{eq:scalar_vertex_uniform} in terms of the out solutions and use the asymptotic behavior of the latter. Within the WKB approach, one has to combine the functions with different signs of the energy when crossing the Stokes line since it is the Stokes phenomenon that gives rise to nonzero particle yield and also governs photon emission and absorption examined in this study (see, e.g., Refs.~\cite{dumlu_prl_2010, dumlu_prd_2011}). It turns out that both $A_{\boldsymbol{n},\lambda}$ and $B_{\boldsymbol{n},\lambda}$ are proportional to $|\alpha_{\boldsymbol{p}}|^2$, where $\alpha_{\boldsymbol{p}}$ is the WKB transition amplitude between the positive-energy state with momentum $\boldsymbol{p}$ and the corresponding negative-energy state (see Appendix~B for more details). Obviously, $|\alpha_{\boldsymbol{p}}|^2$ coincides with $n_{\boldsymbol{p}}$ from Eq.~\eqref{eq:A_spinor_exact} and with $n^{(\text{sc})}_{\boldsymbol{p}}$ from Eq.~\eqref{eq:A_scalar_exact} once the WKB approach is applicable. The coefficients $A_{\boldsymbol{n},\lambda}$ can be calculated quite straightforward and read
\begin{equation}
A^{\text{(WKB)}}_{\boldsymbol{n}, \lambda} = \frac{\alpha}{8\pi^5} \int \! d\boldsymbol{p} \, \frac{(\boldsymbol{p},\boldsymbol{e}_\lambda)^2 p_0^2 (\boldsymbol{P})}{\big [p_0^2 (\boldsymbol{P}) - (\boldsymbol{P},\boldsymbol{n})^2 \big ]^2} \, |\alpha_{\boldsymbol{p}}|^2.
\label{eq:A_scalar_wkb}
\end{equation}
This expression immediately follows from Eq.~\eqref{eq:A_scalar_exact} and it is very accurate in the case of realistic field parameters. For instance, if $E_0 = 0.1E_\text{c}$, then the expression~\eqref{eq:A_scalar_wkb} deviates from Eq.~\eqref{eq:A_scalar_exact} on the level of $1\%$ already for $\tau \gtrsim 30 m^{-1}$. Although one can directly evaluate the exact formula~\eqref{eq:A_scalar_exact} instead of using semiclassical approximations, it is the WKB technique that allows us to obtain a closed-form expression for $B_{\boldsymbol{n},\lambda}$. We arrive at (see Appendix~B)
\begin{equation}
B^{\text{(WKB)}}_{\boldsymbol{n}, \lambda} = \frac{\alpha}{8\pi^5} \int \! d\boldsymbol{p} \, \frac{(\boldsymbol{p},\boldsymbol{e}_\lambda)^2 (\boldsymbol{P},\boldsymbol{n})}{\big [p_0^2 (\boldsymbol{P}) - (\boldsymbol{P},\boldsymbol{n})^2 \big ]^2} \, |\alpha_{\boldsymbol{p}}|^2.
\label{eq:B_scalar_wkb}
\end{equation}
The explicit form of $|\alpha_{\boldsymbol{p}}|^2$ in the domain of interest reads
\begin{equation}
|\alpha_{\boldsymbol{p}}|^2 = \mathrm{e}^{-\pi \tau (\omega_+ + \omega_- + 2eE_0 \tau)},
\label{eq:alpha_scalar_wkb}
\end{equation}
where $\omega_\zeta = \sqrt{m^2 + p_x^2 + p_y^2 + (p_z + \zeta eE_0\tau)^2}$.

Note that the integrands in Eqs.~\eqref{eq:A_scalar_wkb} and \eqref{eq:B_scalar_wkb} contain the difference $p_0^2 (\boldsymbol{P}) - (\boldsymbol{P},\boldsymbol{n})^2$ in the denominator. Since $|eE_0|\tau \gg m$, one has to cancel the term $(p_z + eE_0\tau)^2$ in this difference in order to maximize the coefficients $A_{\boldsymbol{n},\lambda}$ and $B_{\boldsymbol{n},\lambda}$. Accordingly, we will assume $\boldsymbol{n} = \boldsymbol{e}_z$. Furthermore, due to the presence of the factor $(\boldsymbol{P},\boldsymbol{n})$, the expression~\eqref{eq:B_scalar_wkb} vanishes if one chooses $\boldsymbol{n} = \boldsymbol{e}_x$ or $\boldsymbol{n} = \boldsymbol{e}_y$. The polarization vector $\boldsymbol{e}_\lambda$ can now point at any direction in the $xy$ plane as the external background is symmetric, so we choose $\boldsymbol{e}_\lambda = \boldsymbol{e}_x$.

To calculate $A_{zx}$ and $B_{zx}$, one can perform integration in Eqs.~\eqref{eq:A_scalar_wkb} and \eqref{eq:B_scalar_wkb} numerically. Nevertheless, in the realistic regime, one can also evaluate the integrals approximately by means of Laplace's method since the main contribution arises from a small vicinity of $\boldsymbol{p}=0$ due to the exponential suppression of $|\alpha_{\boldsymbol{p}}|^2$. One obtains
\begin{eqnarray}
A_{zx} &\approx& \frac{\alpha}{16 \pi^6} \, m^3 (m\tau)^3 \bigg ( \frac{E_0}{E_\text{c}} \bigg )^{11/2} \mathrm{e}^{-\pi E_\text{c}/E_0}, \label{eq:wkb_Azx_approx}\\
B_{zx} &\approx& -\frac{\alpha}{16 \pi^6} \, m^2 (m\tau)^2 \bigg ( \frac{E_0}{E_\text{c}} \bigg )^{9/2} \mathrm{e}^{-\pi E_\text{c}/E_0}. \label{eq:wkb_Bzx_approx}
\end{eqnarray}
Although these formulas contain the exponential factor which arises in the quantitative analysis of the Sauter-Schwinger mechanism, the factors $\tau^3$ and $\tau^2$ substantially enhance the coefficients. Note that measuring photons emitted along the $y$ axis is a completely unfavorable scenario since $A_{yx} \approx (m/|eE_0\tau|)^4 A_{zx} = \gamma^4 A_{zx} \ll A_{zx}$. Accordingly, for more realistic field parameters, the process of vacuum photon emission predicts a huge number of photons traveling parallel to the electric field. This means that almost isotropic radiation revealed in Ref.~\cite{otto_prd_2017} can be observed only outside the domain considered here. Finally, we note that the coefficient~\eqref{eq:wkb_Bzx_approx} changes its sign if the probe photon travels in the opposite direction.

\begin{figure}[t]
\center{\includegraphics[width=0.98\linewidth]{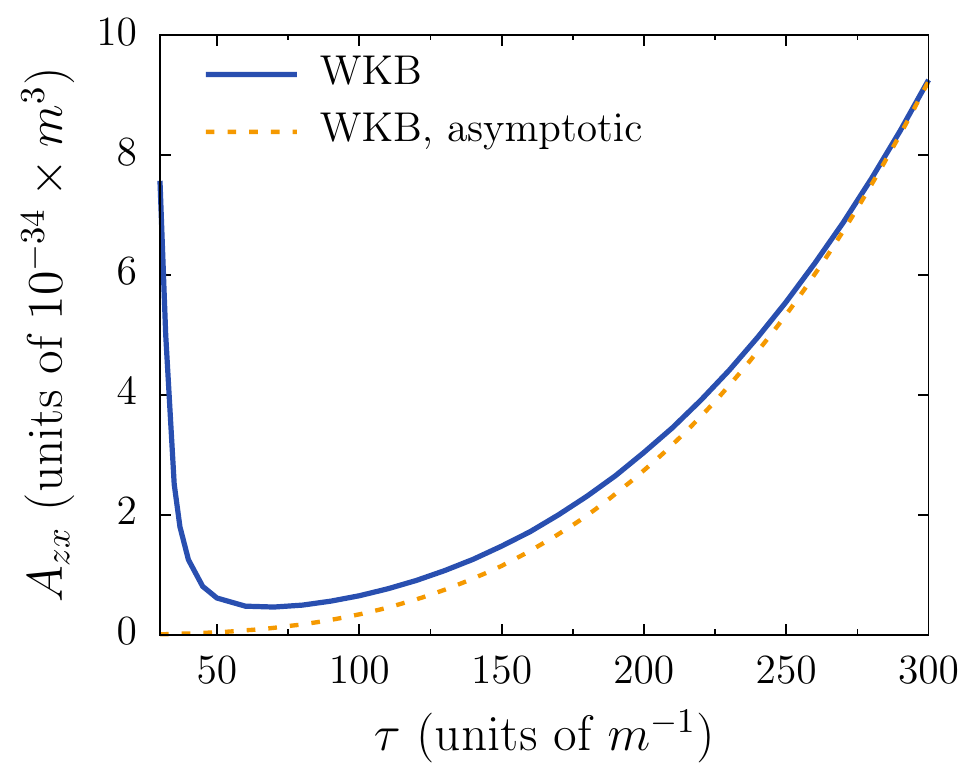}}
\caption{Coefficient $A_{zx}$ as a function of the pulse duration $\tau$ in the case of scalar QED for $E_0 = 0.05E_\text{c}$. The asymptotic behavior (dashed line) is given by Eq.~\eqref{eq:wkb_Azx_approx}.}
\label{fig:res_wkb_A}
\end{figure}

As an illustration, we present the coefficient $A_{zx}$ as a function of $\tau$ for $E_0 = 0.05E_\text{c}$ (see Fig.~\ref{fig:res_wkb_A}). One observes that the approximate expression~\eqref{eq:wkb_Azx_approx}, which predicts the scaling $A_{zx} \sim \tau^3$, is in good agreement with the full WKB calculation.

In what follows, we will examine the experimental scenarios based on measuring either vacuum photon emission or changes in the optical probe beam intensity. We will employ Eqs.~\eqref{eq:wkb_Azx_approx} and \eqref{eq:wkb_Bzx_approx} multiplied by a factor of 2. Both these proposals will also be compared with a direct observation of pairs produced, i.e., the Sauter-Schwinger effect itself. To estimate the total particle yield, one can integrate $|\alpha_{\boldsymbol{p}}|^2$ over $\boldsymbol{p}$ taking into account the factor $V/(2\pi)^3$. This integration can also be carried out by means of Laplace's method. One finds
\begin{equation}
N^{(\text{pairs})} \approx \frac{1}{(2\pi)^3} \, (m^3V)(m\tau) \bigg ( \frac{E_0}{E_\text{c}} \bigg )^{5/2}  \mathrm{e}^{-\pi E_\text{c}/E_0}. \label{eq:wkb_pairs}
\end{equation}

\section{Discussion and experimental prospects} \label{sec:exp}

Having evaluated the coefficients $A_{\boldsymbol{n},\lambda}$ and $B_{\boldsymbol{n},\lambda}$, we can now assess the experimental feasibility of the three following scenarios: (a) direct observation of the Sauter-Schwinger pairs, (b) measuring vacuum radiation, (c) measuring changes in the probe beam intensity. Let us discuss each of these in more detail.

\subsection{Pair production}

The total particle yield can be obtained by means of Eq.~\eqref{eq:wkb_pairs}. Let us introduce a laser wavelength and approximate it by $\lambda \approx 2 \tau$. Assuming that the laser radiation is tightly focused, we note that the volume factor $V$ is proportional to $\lambda^3$. However, taking into account the spatial profile of the external field will definitely reduce the number of pairs since the external field does not have a maximal amplitude in the whole interaction region. To estimate the effect of the spatial inhomogeneities, one can examine the {\it local} values of the particle yield since the realistic laser wavelength is very large~\cite{sevostyanov_prd_2021}. Moreover, the pair production process in the nonperturbative regime is mainly governed by the exponential function $\mathrm{exp}[-\pi E_\text{c}/|E(x)|]$. One can easily verify that for a profile $E(x) = E_0 \cos (2\pi x/\lambda)$, this exponential contributes only within the vicinity $|x| \lesssim 0.08 \lambda$ once $E_0 \lesssim 0.1 E_\text{c}$. Accordingly, in what follows, we will assume that the effective interaction volume amounts to $V = (0.1 \lambda)^3$.

Computing now the number of pairs~\eqref{eq:wkb_pairs}, we will identify the threshold value of the field amplitude $E_0$ depending on $\tau$ by the condition $N^{(\text{pairs})} = 10$. Finally, we point out that the expression~\eqref{eq:wkb_pairs} is quite universal with respect to the choice of the temporal profile of the external field. For instance, in the case of an oscillating background with duration $T = \tau$, one obtains a similar expression which differs from Eq.~\eqref{eq:wkb_pairs} only by factor $2^{3/2}/\pi \approx 0.9$ (see, e.g., Refs.~\cite{sevostyanov_prd_2021, popov_jetp_lett_1971, popov_yad_fiz_1974, ringwald_2001, dunne_wang_2006}).

\subsection{Vacuum emission of soft photons}

As was demonstrated above, the major part of the soft photons emitted from vacuum travels parallel to the $z$ axis. The number of photons can be obtained by integrating Eq.~\eqref{eq:A_B}. Here $d\boldsymbol{k} = k_0^2 \sin \theta dk_0 d\theta d\varphi$, so the number of signal photons in this scenario reads
\begin{equation}
N^{\text{(vac)}} = 2\pi V \sum_{\lambda} \int \limits_{0}^{\pi/2} d\theta \sin \theta \int \limits_{k^0_{\text{min}}}^{k^0_{\text{max}}} dk_0 \, \frac{A_{\boldsymbol{n},\lambda}}{k_0},
\label{eq:N_vac}
\end{equation}
where we imply $\boldsymbol{n} = \sin \theta \boldsymbol{e}_x + \cos \theta \boldsymbol{e}_z$ replacing the integration over $\varphi$ with the factor $2\pi$. The upper limit regarding the $\theta$ integration is inessential as the main contribution arises from a small vicinity of $\theta = 0$. For this reason, we also assume $\boldsymbol{e}_\lambda = \boldsymbol{e}_x$ in the expression for $A_{\boldsymbol{n},\lambda}$ and take into account the second polarization by multiplying the results by a factor of 2. This brings us to the following result:
\begin{equation}
N^{\text{(vac)}} \approx \frac{\alpha}{4 \pi^5} \, (m^3V) (m\tau) \ln \bigg ( \frac{k^0_{\text{max}}}{k^0_{\text{min}}} \bigg ) \bigg ( \frac{E_0}{E_\text{c}} \bigg )^{7/2} \mathrm{e}^{-\pi E_\text{c}/E_0}.
\label{eq:N_vac_final}
\end{equation}
The interval $[k^0_{\text{min}}, k^0_{\text{max}}]$, where one measures the signal photons, is assumed to obey $\ln (k^0_{\text{max}}/k^0_{\text{min}}) = \ln 4$, which corresponds to, e.g., a frequency interval twice as large as the full width of the visible spectrum. The threshold of the process is defined via $N^{\text{(vac)}} = 10$, i.e., one has to be able to detect at least ten signal photons. It is already seen that the number of soft photons is suppressed by additional factor $E_0/E_\text{c}$ compared to the particle yield~\eqref{eq:wkb_pairs}. In what follows, we will find out how this affects the experimental prospects of this scenario.

\subsection{Probe beam intensity}

Here we propose to measure a relative change in the intensity of the probe photon beam, which can be evaluated via
\begin{equation}
\eta \equiv \frac{N^{\text{(ph)}}}{N} = \frac{2 (2\pi)^3 |B_{zx}|}{q_0^2},
\label{eq:N_ratio_2}
\end{equation}
where we will employ Eq.~\eqref{eq:wkb_Bzx_approx}. Note that the probe beam is orthogonal to the propagation direction of the lasers that we treat here as a classical background, so the primary laser beams as well as the photons emitted via the tadpole diagram will not obscure the probe quanta. Moreover, the vacuum term~\eqref{eq:N_vac_final} is many orders of magnitude smaller than the initial number of probe photons $N$ even if we sum over all spatial directions.

The ratio~\eqref{eq:N_ratio_2} can be enhanced by choosing a low frequency of the probe photons. We will assume $q_0 = 10^{-6} m$, which corresponds to a wavelength of $2.4~\mu\text{m}$. The realistic threshold of this scenario is set to $\eta = 0.01$, i.e., the relative intensity change should amount to at least one percent.

\begin{figure}[t]
\center{\includegraphics[width=0.98\linewidth]{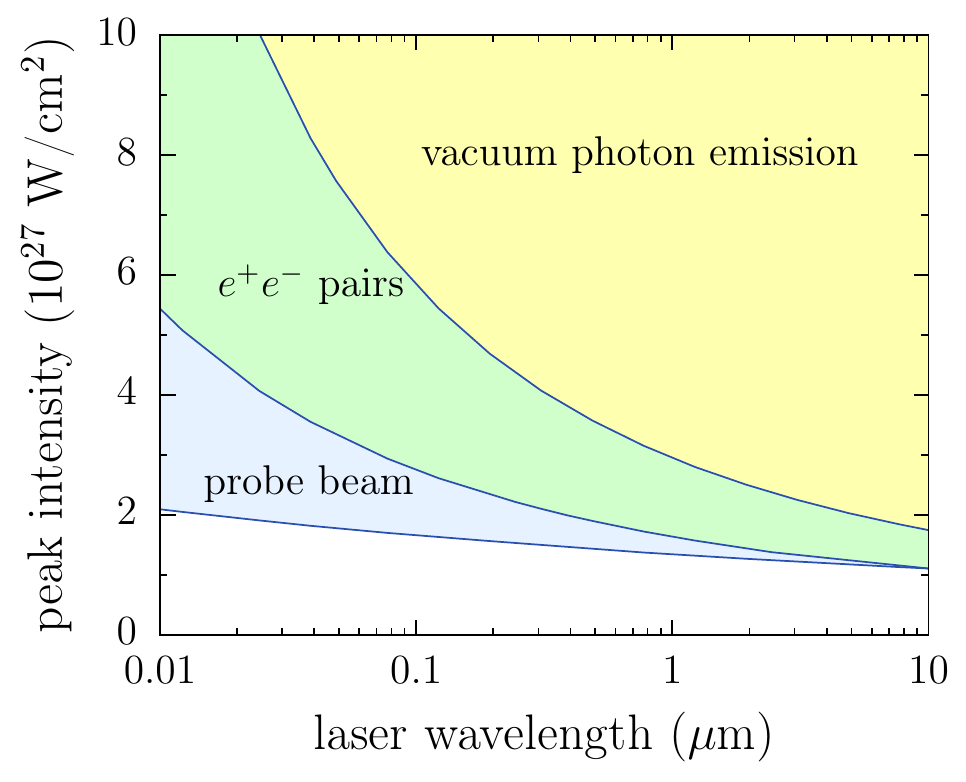}}
\caption{Threshold values of the laser peak intensity that are necessary for measuring vacuum photon emission (upper curve), the Sauter-Schwinger effect itself (middle curve), and changes in the intensity of the probe photon beam (lower curve).}
\label{fig:final}
\end{figure}

\subsection{Comparison}

In Fig.~\ref{fig:final} we present the threshold values of the external field amplitude in terms of the laser intensity for various values of the laser wavelength. The curves were found to be stable with respect to the changes of the parameters involved in the pre-exponential factors and threshold values of $N^{(\text{pairs})}$, $N^{(\text{vac})}$, and $\eta$, respectively. First, one observes that although the peak intensity corresponding to the field strength $E_\text{c}$ is $2.3 \times 10^{29}~\text{W/cm}^2$, the real threshold of the pair production process is about two orders of magnitude lower due to the presence of a large pre-exponential factor (it was examined also, e.g., in Ref.~\cite{bulanov_prl_2010}). Second, it turns out that measuring vacuum photon emission is not that promising compared to a direct detection of the Sauter-Schwinger pairs. Note that both of these contributions are proportional to $\tau^4$, so the threshold values of the laser intensity decrease rather rapidly with increasing $\lambda$. Nevertheless, the probe-photon technique turns out to be more advantageous, especially for smaller values of $\lambda$.

Finally, let us provide several additional remarks concerning the experimental implementation of the three setups. If the external field contains several cycles for a given half period $\tau = \lambda/2$, the particle yield as well as the number of soft photons will be enhanced by the corresponding factor, which will also slightly lower the position of the curves in Fig.~\ref{fig:final}. On the other hand, it is crucial to have a unipolar laser background to achieve notable changes in the intensity of the probe photon beam. Further developments in the practical generation of such unipolar pulses (see Ref.~\cite{kozlov_pra_2011}), especially in the domain $\lambda \lesssim 0.1~\mu\text{m}$ should make the probe-beam technique an efficient tool for measuring nonperturbative strong-field QED effects.

\section{Conclusion} \label{sec:conclusions}

In this study, we computed the number density of soft photons emitted in the presence of a strong electric background. The main goals were (a) to examine the role of additional (probe) photons in the initial state and (b) to assess the experimental prospects of two proposals with regard to measuring electromagnetic radiation instead of detecting electron-positron pairs. To perform the necessary computations, we employed the Furry picture formalism together with the WKB approach allowing one to investigate a realistic regime of the field parameters. It was demonstrated that the technique which was proposed here and is based on using an additional optical probe photon beam can be utilized in order to lower the pair production threshold although the pure vacuum radiation is unlikely to be detected prior to the onset of the Sauter-Schwinger effect.

\begin{acknowledgments}
This work was supported by Russian Foundation for Basic Research (RFBR) and Deutsche Forschungsgemeinschaft (DFG) (Grants No. 20-52-12017 and No. PL 254/10-1). I.A.A. also acknowledges the support from the Foundation for the advancement of theoretical physics and mathematics ``BASIS'' and from the German-Russian Interdisciplinary Science Center (G-RISC) funded by the German Federal Foreign Office via the German Academic Exchange Service (DAAD).
\end{acknowledgments}

\appendix

\section{Calculation of $A_{\boldsymbol{n},\lambda}$ in spinor QED} \label{sec:appendix_A_spinor}

Let us first represent the expression~\eqref{eq:ph_number_density_uniform} in the following form:
\begin{equation}
\frac{(2\pi)^3}{V} \, n_{\boldsymbol{k}, \lambda}^{\text{(vac)}} = \frac{\alpha}{4\pi^2} \frac{1}{k_0} \sum_{s, s'} \int \! d\boldsymbol{p} \, \bigg | \frac{a_{s,s'}}{k_0} + b_{s,s'} + ... \bigg |^2.
\label{eq:appA_a_b}
\end{equation}
Then the coefficient $A_{\boldsymbol{n}, \lambda}$ reads
\begin{equation}
A_{\boldsymbol{n}, \lambda} = \frac{\alpha}{32\pi^5} \sum_{s, s'} \int \! d\boldsymbol{p} \, | a_{s,s'} |^2.
\label{eq:appA_A}
\end{equation}
As was stated in the text, the leading contribution $a_{s,s'}/k_0$ arises from integrating over $t \in [t_\text{out}, +\infty)$ in Eq.~\eqref{eq:ph_number_density_uniform}. To perform this integration, one has to express the in solutions in terms of the out ones and make use of the asymptotic behavior of the latter. Here we will need the following scalar products:
\begin{equation}
G \big ( {}_\zeta \big | {}^\kappa \big )_{\boldsymbol{p},s,\boldsymbol{p}',s'} = ({}_\zeta \varphi_{\boldsymbol{p},s}, {}^\kappa \varphi_{\boldsymbol{p}',s'} ),
\label{eq:appA_G}
\end{equation}
where $\zeta, \kappa = \pm$. In the case of a spatially homogeneous external background, one finds
\begin{equation}
G \big ( {}_\zeta \big | {}^\kappa \big )_{\boldsymbol{p},s,\boldsymbol{p}',s'} = \delta_{s,s'} \delta (\zeta \boldsymbol{p} - \kappa \boldsymbol{p}') g \big ( {}_\zeta \big | {}^\kappa \big )_{\boldsymbol{p},s}.
\label{eq:appA_g}
\end{equation}
Integrating over $t \in [t_\text{out}, +\infty)$, one obtains
\begin{eqnarray}
a_{s,s'} &=& \frac{i(\bar{u}_{\boldsymbol{P},s} \gamma^\mu \varepsilon_\mu^* u_{\boldsymbol{P},s'})}{1 - (\boldsymbol{P}, \boldsymbol{n})/p_0(\boldsymbol{P})} \, g \big ( {}_+ \big | {}^+ \big )_{\boldsymbol{p},s} \, g^* \big ( {}_- \big | {}^+ \big )_{-\boldsymbol{p},s'} \nonumber \\
{} &+& \frac{i(\bar{v}_{\boldsymbol{P},s} \gamma^\mu \varepsilon_\mu^* v_{\boldsymbol{P},s'})}{1 + (\boldsymbol{P}, \boldsymbol{n})/p_0(\boldsymbol{P})} \, g \big ( {}_+ \big | {}^- \big )_{\boldsymbol{p},s} \, g^* \big ( {}_- \big | {}^- \big )_{-\boldsymbol{p},s'},
\label{eq:appA_ag}
\end{eqnarray}
where $\boldsymbol{P} = \boldsymbol{p} - e\mathcal{A}_0$ and $u_{\boldsymbol{p},s}$ ($v_{\boldsymbol{p},s}$) are constant bispinors corresponding to the positive (negative) energy solutions of the free Dirac equation [cf.~Eq.~(B2) in Ref.~\cite{otto_prd_2017}]. Taking the mod-square of Eq.~\eqref{eq:appA_ag}, one has to compute standard traces involving bispinors and the gamma matrices. The crucial point here is that in the case of a Sauter profile~\eqref{eq:ext_field_sauter}, the $g$ coefficients are known explicitly~\cite{fradkin_gitman_shvartsman}. The coefficients with opposite signs yields the number density $n_{\boldsymbol{p}}$ of the electrons (positrons) produced:
\begin{equation}
n_{\boldsymbol{p}} = \big | g \big ( {}_+ \big | {}^- \big )_{\boldsymbol{p},s} \big |^2 = \big | g \big ( {}_- \big | {}^+ \big )_{-\boldsymbol{p},s} \big |^2.
\end{equation}
The coefficients with only positive (negative) signs obey
\begin{equation}
\big | g \big ( {}_+ \big | {}^+ \big )_{\boldsymbol{p},s} \big |^2 = \big | g \big ( {}_- \big | {}^- \big )_{-\boldsymbol{p},s} \big |^2 = 1 - n_{\boldsymbol{p}}.
\end{equation}
Then straightforward calculations bring us to Eq.~\eqref{eq:A_spinor_exact}.

\section{WKB calculation of $B_{\boldsymbol{n},\lambda}$} \label{sec:appendix_B_wkb}

As was done in Appendix~A, it is again convenient to represent the photon number density~\eqref{eq:scalar_vertex_uniform} in the following form:
\begin{equation}
\frac{(2\pi)^3}{V} \, n_{\boldsymbol{k}, \lambda}^{\text{(sc)}} = \frac{\alpha}{\pi^2} \frac{1}{k^0} \int \! d\boldsymbol{p} \, (\boldsymbol{p},\boldsymbol{e}_\lambda)^2 \bigg | \frac{a}{k_0} + b + ... \bigg |^2.
\label{eq:appB_a_b}
\end{equation}
Performing calculations similar to those outline in Appendix~A, one can first derive the expression~\eqref{eq:A_scalar_exact}. In fact, these computations are easier as one deals only with scalar functions instead of matrices. The coefficient $B_{\boldsymbol{n}, \lambda}$ should be evaluated via
\begin{equation}
B_{\boldsymbol{n}, \lambda} = \frac{\alpha}{4\pi^5} \int \! d\boldsymbol{p} \, (\boldsymbol{p},\boldsymbol{e}_\lambda)^2 \mathrm{Re} (a^* b).
\label{eq:appB_B}
\end{equation}
Within the WKB approximation, one can explicitly construct the solutions taking into account the Stokes phenomenon and then find $a$ and $b$. One obtains
\begin{equation}
a = \frac{i}{p_0} \frac{\alpha_{\boldsymbol{p}}}{1 - (\boldsymbol{p}, \boldsymbol{n})^2/p_0^2},
\label{eq:appB_a}
\end{equation}
where
\begin{eqnarray}
\alpha_{\boldsymbol{p}} &=& - i  \, \mathrm{exp} \bigg [ 2 i \int \limits_{t_\text{in}}^{t_{\boldsymbol{p}}} p_0 (\boldsymbol{p} - e\boldsymbol{\mathcal{A}}) dt \bigg ] \, \frac{\cosh (\pi \tau \beta_{\boldsymbol{p}} /2)}{\sinh (\pi \tau \gamma_{\boldsymbol{p}} )}, \label{eq:appB_alpha} \\
\beta_{\boldsymbol{p}} &=& (\omega_+ - \omega_-) \, \mathrm{sgn}~p_z - 2eE_0 \tau, \\
\gamma_{\boldsymbol{p}} &=& \sqrt{m^2+p_x^2+p_y^2 + (|p_z| + eE_0 \tau)^2}.
\end{eqnarray}
Here $t_{\boldsymbol{p}}$ is the time instant where the Stokes line intersects the real axis. Decomposing the amplitude of the process with respect to $k_0$, one finds that the term $b$ in Eq.~\eqref{eq:appB_a_b} contains numerous contributions. However, the real part in Eq.~\eqref{eq:appB_B} leaves only one of them given in Eq.~\eqref{eq:B_scalar_wkb}. In the semiclassical approximation, the absolute value of Eq.~\eqref{eq:appB_alpha} coincides with Eq.~\eqref{eq:alpha_scalar_wkb}.


\end{document}